\documentclass[%
 reprint,
 superscriptaddress,
 twocolumn,
 amsmath,amssymb,
 aps,
pra,
]{revtex4-1}
\usepackage[utf8]{inputenc}

\usepackage[paperwidth=210mm,paperheight=297mm,centering,hmargin=2.3cm,tmargin=2.8cm,bmargin=3.4cm]{geometry}
\usepackage[normalem]{ulem}
\usepackage{courier}
\usepackage{graphicx,epsf}
\usepackage{amsfonts}
\usepackage{amssymb}
\usepackage{physics}
\usepackage{nth}
\usepackage{amsmath}
\usepackage{overpic}
\usepackage{fancyhdr}
\usepackage{hyperref}
\usepackage{xcolor}
\hypersetup{
	colorlinks   = true, 
	urlcolor     = blue, 
	linkcolor    = red, 
	citecolor   = blue 
}

\newcommand{\half}[1][1]{\frac{#1}{2}}
\newcommand{\inv}[1]{\frac{1}{#1}}
\newcommand{\ad}{\hat{a}^{\dag}}
\newcommand{\Ad}{A^{\dag}}
\newcommand{\ir}{\mathrm{i}}
\newcommand{\er}{\mathrm{e}}
\newcommand{\gc}{g_{\mathrm{c}}}
\newcommand{\gcdyn}{g_{\mathrm{c}}^{\mathrm{dyn}}}
\newcommand{\gf}{g_{\mathrm{f}}}
\newcommand{\muc}{\mu_{\mathrm{c}}}
\newcommand{\muf}{\mu_{\mathrm{f}}}
\newcommand{\omi}{\omega_{0,k}}
\newcommand{\omf}{\omega_{\mathrm{f},k}}
\newcommand{\omfz}{\omega_{\mathrm{f},0}}
\newcommand{\gap}{\Delta_{\mathrm{f}}}

\newcommand{\idk}{\int\!\frac{\dd{k}}{2\pi}\,}
\newcommand{\dk}{\frac{\dd{k}}{2\pi}\,}

\begin{document}
\title{Dynamical quantum phase transition in a bosonic system with long-range interactions}

\author{Marvin Syed}
\email{syed@thphys.uni-heidelberg.de}
\affiliation{Institut f\"ur Theoretische Physik, Universit\"at 
Heidelberg, D-69120 Heidelberg, Germany}
\author{Tilman Enss}
\affiliation{Institut f\"ur Theoretische Physik, Universit\"at 
Heidelberg, D-69120 Heidelberg, Germany}
\author{Nicol\`o Defenu}
\affiliation{Institute for Theoretical Physics, ETH Z$\ddot{u}$rich, Wolfgang-Pauli-Str.\ 27, 8093 Z$\ddot{u}$rich, Switzerland}
\affiliation{Institut f\"ur Theoretische Physik, Universit\"at 
Heidelberg, D-69120 Heidelberg, Germany}

\begin{abstract}
	In this paper, we investigate the dynamical quantum phase transitions appearing in the Loschmidt echo and the time-dependent order parameter of a quantum system of harmonically coupled degenerate bosons as a function of the power-law decay $\sigma$ of long-range interactions.         
	Following a sudden quench, the nonequilibrium dynamics of this system are governed by a set of nonlinear coupled Ermakov equations.
	To solve them, we develop an analytical approximation valid at late times.
	Based on this approximation, we show that the emergence of a dynamical quantum phase transition hinges on the generation of a finite mass gap following the quench, starting from a massless initial state. 
	In general, we can define two distinct dynamical phases characterized by the finiteness of the post-quench mass gap.
	The Loschmidt echo exhibits periodical nonanalytic cusps whenever the initial state has a vanishing mass gap and the final state has a finite mass gap. These cusps are shown to coincide with the maxima of the time-dependent long-range correlations.
\end{abstract}

\maketitle

\section{Introduction}
Recent experimental advances, especially in cold atoms\,\cite{gring2012relaxation, langen2013local, hild2014far, bordia2016coupling, bordia2017periodically, bordia2017probing,muniz2020exploring} and trapped ions\,\cite{monroe2019programmable}, made the investigation of the quantum dynamics of many-body systems feasible and aroused great interest in the theoretical characterization of dynamical critical phenomena\, \cite{polkovnikov_colloquium_2011}. These phenomena are related to the existence of dynamical phases of matter and of the associated dynamical quantum phase transitions (DQPTs)\,\cite{heyl_dynamical_2013,heyl_dynamical_2019}.

DQPTs were first characterized in analogy to Landau theory by studying the behavior of dynamical order parameters\,\cite{yuzbashyan2006relaxation,barmettler2009relaxation,eckstein2009thermalization,sciolla2010quantum,mitra2012time}. More recently a second characterization appeared, based on the appearance of nonanalytic cusps in the Loschmidt echo rate function\,\cite{heyl_dynamical_2013,heyl_dynamical_2019}. The existence of such nonanalyticities has been traced back to the analogy between the  amplitude $\braket{\Psi_{0}}{\Psi(t)}\equiv \bra{\Psi_{0}}\exp(-\ir Ht) \ket{\Psi_{0}}$ of the initial and time-evolved states and the classical partition function $Z(\beta)=\tr\exp(-\beta H)$. Theoretical evidence of DQPTs in the Loschmidt echo return rate was found in numerous quantum systems\,\cite{heyl_dynamical_2013, heyl_dynamical_2014, halimeh_dynamical_2017, vajna_disentangling_2014, vajna_topological_2015, pozsgay_dynamical_2013, piroli_non-analytic_2018, schmitt_dynamical_2015, campbell_criticality_2016, zunkovic_dynamical_2016, weidinger_dynamical_2017, lang2018dynamical, abdi_dynamical_2019} and their connection with the singular dynamics of the order parameter has been explicitly analyzed in the Ising model\,\cite{zunkovic_dynamical_2018}.
	
Experimental evidence of DQPTs in many-body quantum systems was mainly confined to quantum spin chains, which can be simulated in trapped ion systems and display extended nonlocal interactions\,\cite{jurcevic2017observation,zhang2017observation}. It is, therefore, not surprising that theoretical studies of DQPTs in systems with power-law decaying interactions have been thriving, both on spins\,\cite{zunkovic_dynamical_2016,zunkovic_dynamical_2018,halimeh_dynamical_2017, lang2018concurrence,homrighausen2017anomalous,halimeh2020quasiparticle} and Fermi\,\cite{vajna_topological_2015,dutta_probing_2017,defenu2019dynamical,uhrich2020out} systems.

Despite the wide range of investigations, several important questions regarding the critical dynamics of quantum models remain open. In particular, the relation between the occurrence of dynamical quantum phase transitions and the quasiparticle spectrum in systems with bosonic excitations has been rather limited, while nonanalytic kinetic terms due to long-range interactions are known to produce several anomalous dynamical phases in spin and Fermi systems\,\cite{halimeh2020quasiparticle,defenu2019dynamical,defenu2019universal,uhrich2020out}. In this paper we extend the study of dynamical phase transitions into the bosonic realm, showing that the nonanalytic momentum terms in the quasiparticle dispersion relation do not produce additional critical points with respect to the nearest-neighbour case. Conversely, the nonanalytic dispersion relation is shown to push the cusps signaling the DQPT to higher derivative orders in the  Loschmidt echo.     

The focus of our investigations is the quantum extension of the Spherical model introduced by Berlin and Kac\,\cite{berlin_spherical_1952}, which describes a system of harmonically coupled quantum oscillators with a global constraint imposed on the expectation value of their positions, namely $\sum_{i=1}^{N} \langle \hat{s}_{i}^{2}\rangle=N/4$\,\cite{vojta_quantum_1996}. 
Despite its quadratic nature, the presence of the global spherical constraint induces a quantum critical point in this model, depending on the number of spatial dimensions and on the harmonic interaction shape. It is worth noting that in the classical limit, the spherical model's free energy corresponds to the one of $O(n)$-symmetric spin systems in the $n \rightarrow \infty$ limit\,\cite{stanley_spherical_1968}.
Therefore, the scientific interest in the spherical model is primarily due to its universal behavior, which qualitatively describes any quantum critical point where a continuous symmetry is spontaneously broken. Nevertheless, a concrete experimental realisation of the spherical model has been proposed using multidimensional laser mode lattices\,\cite{schwartz_laser_2013}.
	
When considering universal properties, this correspondence also extends to continuous $O(n)$ field theories, which in the $n\to\infty$ limit lie in the same universality class of the spherical model\,\cite{joyce1966spherical,defenu2015fixed}. At equilibrium, the correspondence between quantum critical points and classical phase transitions\,\cite{sachdev2011quantum} allows one to conclude the existence of a critical point also in the quantum case\,\cite{defenu2017criticality}.  Therefore, apart from the interest in the study of bosonic many-body quantum systems with long-range interactions, our studies also target important aspects of universality in DQPTs. Indeed, in equilibrium it is possible to relate the universal behavior of $d$-dimensional long-range models with decay exponent $d+\sigma$ to that of the corresponding nearest-neighbour system in dimension $d_{\mathrm{eff}}=2d/\sigma$, constituting a superuniversal scaling relation\,\cite{defenu2020}.  Even if the concept of universality does not fully apply to the dynamical realm\,\cite{heyl2015scaling},  we employ the effective dimension relation to show that our findings are consistent with the ones obtained for $O(n)$ models at large $n$ in Refs.\,\cite{weidinger_dynamical_2017,chandran_equilibration_2013} and provide evidence of universality also in DQPTs.

The paper proceeds as follows: In Sec.\,\ref{sec:model} we will introduce the spherical model and briefly discuss its equilibrium properties. Then, in Sec.\,\ref{subsec:dyn_case} we will consider the time-dependent case, focusing on the case of an instantaneous quench. An exact solution to the spherical model dynamics will be obtained by a time-dependent canonical transformation\,\cite{lewis_classical_1967,sheng_quantum_1995}, leading to a system of Ermakov equations coupled by the global constraint.
In Sec.\,\ref{sec:results} an approximate solution to the coupled Ermakov equations will be developed, i.e., the \emph{step approximation}. The validity of such an approximate solution will be corroborated by comparing it to a numerical solution of the differential equations. 

\begin{figure*}
\centering
\includegraphics[width=0.9\linewidth]{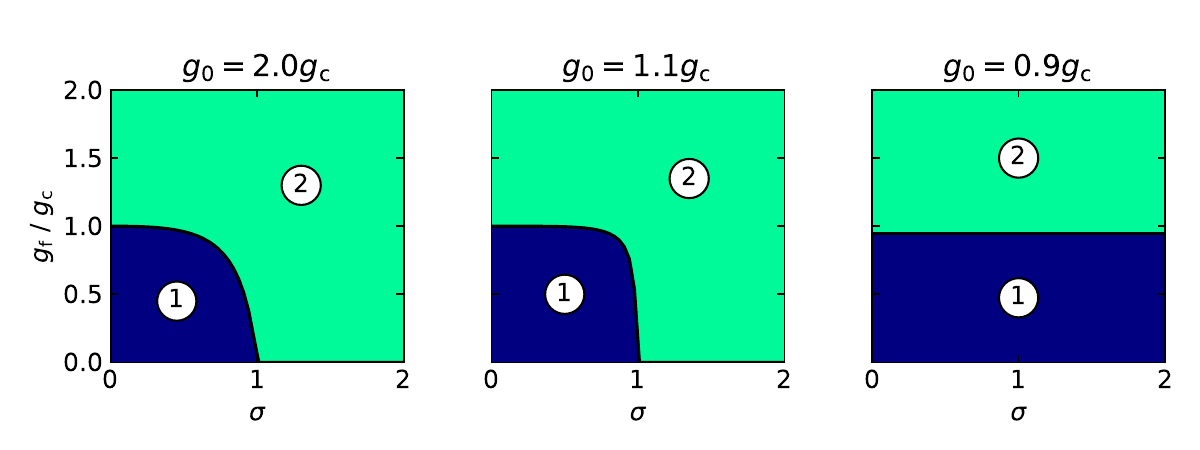}
\caption{Dynamical phase diagram of the spherical model with power-law decay $\sigma$ for quenches from different initial couplings $g_0$ to final coupling $g_f$. We identify two different phases based on the long-time limit of the post-quench gap which can either vanish or become nonzero. The gapless (or ordered) dynamical phase (1) has a vanishing post-quench gap and allows for massless excitations, while the gapped (or disordered) dynamical phase (2) has a nonzero post-quench gap and massive excitations. The phase diagram strongly depends on $g_0$, and there are two main cases: For $g_0>\gc$ (strong initial quantum fluctuations) there is no ordered dynamical phase if $\sigma>1$ (left and middle plot). For $g_0 \leq \gc$ (weaker quantum fluctuations), there is always an ordered dynamical phase for $\sigma<2$.}
\label{fig:dynphase}
\end{figure*}

Having justified the \emph{step approximation}, we will employ it to obtain the dynamical phase diagram for the long-range spherical model, which is displayed in Fig.\,\ref{fig:dynphase}.
Additionally, analytical expressions for the Loschmidt echo and the long-range correlations will be derived and numerically evaluated. The resulting return rate function exhibits periodically spaced nonanalytic cusps signaling the presence of a dynamical quantum phase transition.
The conclusions of our investigations will be discussed in Sec.\,\ref{sec:end}, where we summarize the results and give an outlook to future directions of research.
  	
\section{The model}
\label{sec:model}
\subsection{Equilibrium}
In the absence of any symmetry breaking field, the Hamiltonian of the spherical model reads
\begin{align}\label{eq:ham_eq}
H = \frac{g}{2} \sum_i \hat{p}_i^2 + \frac{1}{2} \sum_{i,j} U_{ij} \hat{s}_i \hat{s}_j
	+ \mu \left( \sum_i \hat{s}_i^2 - \frac{N}{4} \right) \ ,
\end{align}
where the $\hat{s}_i$ and $\hat{p}_i$ are canonically conjugate
hermitian operators on a one-dimensional lattice, such that
$\comm{\hat{s}_i}{\hat{p}_j} = \ir \delta_{i,j}$ (with $\hbar=1$). The
coupling $g$ regulates the strength of quantum fluctuations; in the limit $g \rightarrow 0$ the Hamiltonian in Eq.\,\eqref{eq:ham_eq} reduces to the one of the classical spherical model\,\cite{joyce1966spherical}. The spherical constraint 
\begin{align}
\label{eq:constraint}
\ev{\sum_i \hat{s}_i^2} = \frac{N}{4} 
\end{align}
is enforced by a Lagrange multiplier $\mu$. As mentioned in the introduction, we are going to consider long-range power-law decaying couplings of the form
\begin{align}
U_{ij} = -\frac{1}{\abs{ i-j }^{\sigma + 1}} \quad ,
\end{align}
where the lattice spacing has been set to unity for simplicity.
In one dimension the total energy of the system is then only extensive for $\sigma>0$ \cite{campa_statistical_2009}.

It is convenient to recast the Hamiltonian in Eq.\,\eqref{eq:ham_eq} in Fourier space, yielding
\begin{align}\label{eq:qho}
	H = \frac{g}{2} \sum_k \hat{p}_k \hat{p}_{-k} + 
		\frac{1}{2g} \sum_k \omega_k^2 \hat{s}_k \hat{s}_{-k} 
\end{align}
where the frequency reads
\begin{align}\label{eq:disp}
	\omega_k^2 = 2g(\mu + U_k / 2) \quad ,
\end{align}
and a constant term in the Hamiltonian has been neglected.
The Fourier transformed interaction $U_k$ approaches the closed form
\begin{align}
	U_k = -2 \Re\qty[\operatorname{Li}_{\sigma+1}\qty(\er^{\ir k})]
\end{align}
in the thermodynamic limit, where $\operatorname{Li}_s(z)$ denotes the polylogarithm.
In Fourier space the Hamiltonian describes $N$ uncoupled harmonic oscillators and may be recast in the diagonal form
\begin{align}
	H = \sum_k \omega_k \left( \ad_k \hat{a}_k + \frac{1}{2} \right)
\end{align}
by means of the well-known transformations
\begin{subequations}
	\begin{align}
		\hat{a}_k   &= \sqrt{\frac{\omega_k}{2g}} \hat{s}_k + 
			\ir\sqrt{\frac{g}{2\omega_k}} \hat{p}_{-k} \quad , \\
		\ad_k &= \sqrt{\frac{\omega_k}{2g}} \hat{s}_{-k} - 
			\ir\sqrt{\frac{g}{2\omega_k}} \hat{p}_{k}  \quad ,
	\end{align}
\end{subequations}
ensuring that the commutation relation $\comm*{\hat{a}_k}{\ad_{k'}} = \delta_{kk'}$ holds.
Evaluating the spherical constraint \eqref{eq:constraint} in the ground state $\ket{0} = \prod_k \ket{k,0}$, one obtains
\begin{align}\label{eq:sph}
	0 = -\frac{1}{4} + \frac{g}{N} \sum_k \frac{1}{2\omega_k} \quad .
\end{align}
The existence of a quantum phase transition can be connected with the appearance of a gapless point in the excitation spectrum of the system. Since the Hamiltonian in Eq.\,\eqref{eq:ham_eq} is homogeneous and isotropic one expects such a soft mode to appear at zero momentum ($k=0$), leading to a homogeneous order parameter. 

Indeed, for a thermodynamic system $N\to\infty$ the summation in Eq.\,\eqref{eq:sph} may be cast into a continuous integral, which would have to grow indefinitely in the limit $g\to 0$ in order for the constraint in Eq.\,\eqref{eq:sph} to be satisfied. Then, as long as the integral in Eq.\,\eqref{eq:sph} converges for all $\mu$, there must exist a critical value $g_{c}$ below which the continuous approximation fails and the system ground state changes. At $g=g_{c}$ the maximum of the integral is attained and the parameter  $\mu$ tends to the critical value $\muc = -U_{k=0}/2$, where the dispersion relation in Eq.\,\eqref{eq:disp} becomes gapless. Therefore,   the critical coupling $\gc$ is given by the equation
\begin{align}\label{eq:gc}
0 = -\frac{1}{4} + \frac{\sqrt{\gc}}{2N} \sum_k \frac{1}{\sqrt{2\muc + U_k}} \quad .
\end{align}
Since $U_k - U_{k=0} \propto \abs{k}^\sigma$ for small $k$, a quantum critical point (QCP) can only exist for $\sigma < 2$. The same procedure can be applied to locate the finite temperature phase transition, once the proper thermal occupation is included in Eq.\,\eqref{eq:sph}, leading to the threshold $\sigma<1$ for the existence of the classical phase transition\,\cite{vojta_quantum_1996}.
For the rest of the paper, we will work mostly in the region $0 < \sigma < 2$, where the energy still scales extensively and the equilibrium phase transition is possible.

\subsection{The dynamical case}
\label{subsec:dyn_case}
Our study focuses on the dynamical behavior of the spherical model after a sudden change of the coupling $g$. It will be convenient to first derive the general time-dependent solution of the model and then focus on the quench case. Since the spherical model is described by the quadratic Hamiltonian in Eq.\,\eqref{eq:qho} its dynamical behavior can be obtained by the study of the harmonic oscillators with time-dependent mass $1/g(t)$. This problem can be conveniently described in the time-dependent canonical transformation formalism\,\cite{sheng_quantum_1995}. We first introduce the generating function
\begin{align}
\begin{split}
	F(\hat{s}_k, & \hat{P}_k, t) = \\ & \sum_k \biggl[\frac{\hat{s}_k \hat{P}_{-k} + \hat{P}_k \hat{s}_{-k}}{2\xi_k(t)} -\frac{\Phi_k(t)}{2\xi_k(t)} \hat{s}_k \hat{s}_{-k} \biggr] \ ,
\end{split}
\end{align}
which defines the transformed coordinates and momenta via the relations $\hat{p}_k=\partial F/\partial \hat{s}_{-k}$ and $\hat{S}_k=\partial F/\partial \hat{P}_{-k}$. Accordingly, the transformed position and momentum operators read
\begin{subequations}
\begin{align}
\hat{S}_k &= \frac{\hat{s}_k}{\xi_k(t)} \\
\hat{P}_k &= \xi_k(t) \hat{p}_k + \Phi_k(t) \hat{s}_k\quad,
\end{align}
\end{subequations}
where the functions $\xi(t)$ and $\Phi(t)$ have to be chosen in such a way that the problem is reduced to an effective time-\emph{independent} one. A convenient choice reads
\begin{subequations}
\begin{align}
g(t) \Phi_k(t) + \dot{\xi}_k(t) &= 0 \qq{and} \label{eq:erm1} \\
\omega_k^2(t) \xi_k(t) - g(t) \dot{\Phi}_k &= \frac{\lambda_k^2 g^2(t)}{\xi_k^3(t)} \label{eq:erm2}
\end{align}
\end{subequations}
with arbitrary time independent coefficients $\lambda_k$, which have to be chosen in order to satisfy the initial conditions of the dynamics. 

Equations \eqref{eq:erm1} and\,\eqref{eq:erm2} may be rephrased in terms of the Ermakov equation
\begin{align}\label{eq:erm}
\ddot{\xi}_k(t) + \gamma(t) \dot{\xi}_k(t) + 
\omega_k^2(t) \xi_k(t) = \frac{\lambda_k^2 g^2(t)}{\xi_k^3(t)}
\end{align}
with the initial conditions
\begin{align}
\xi_k(0) = 1 \qq{and} \Phi_k(0) = \dot{\xi}_k(t) = 0 \quad .
\end{align}
The damping term $\gamma(t) = -\dv*{\log g(t)}{t} = 0$ for $t>0$ in case of a sudden quench at $t=0$.
The Hamiltonian for the transformed variables follows from the relation $H' = H + \pdv*{F}{t}$, yielding
\begin{align}
H' = \sum_k \frac{g(t)}{2\xi_k^2(t)} \left( \hat{P}_k \hat{P}_{-k} + \lambda_k^2 \hat{S}_k \hat{S}_{-k} \right) \quad,
\end{align}
which may be recast as a time-independent problem in terms of the effective time 
\begin{align}
	\tau=\int_{0}^{t} \frac{g(t')}{2\xi_{k}(t')^{2}}\dd{t'} \quad .
\end{align} 		
		
The transformed Hamiltonian can again be diagonalized using a proper definition of ladder operators (referring to it as $H$ henceforth),
\begin{align}
H = \sum_k \frac{g(t)}{\xi_k^2(t)} \lambda_k \left(\hat{A}^{\dagger}_k \hat{A}_k + \frac{1}{2}\right) \quad , 
\end{align}
where 
\begin{subequations}
\begin{align}
\hat{A}_k &= \frac{1}{\sqrt{2\lambda_k}} (\lambda_k^2 \hat{S}_k + \ir \hat{P}_{-k}) \quad , \\
\hat{A}^{\dagger}_k &= \frac{1}{\sqrt{2\lambda_k}} (\lambda_k^2 \hat{S}_{-k} - 
\ir \hat{P}_{k}) \quad .
\end{align}
\end{subequations}

However, products of eigenstates $\ket{k,n}$ of $\Ad_k A_k$ are not eigenstates of the Hamiltonian anymore because of the factor $g(t) / \xi_k^2(t)$. Instead, the eigenstates are given by products of the states
		\begin{align}
			\begin{split}
				\ket{\psi_{k,n_k}(t)} & = \\ \exp\biggl[ 
					-\ir & \lambda_k \left( n_k + \frac{1}{2} \right) \int_0^t
						\frac{\dd{t'} g(t')}{\xi_k^2(t')}
				\biggr] \ket{k,n_k}
			\end{split}
		\end{align}
up to an additional time-dependent phase. Evaluating the constraint \eqref{eq:constraint} in the time-dependent ground state $\ket{\Psi_0(t)} = \prod_k \ket{\psi_{k, 0}(t)}$, one obtains the dynamical constraint equation
		\begin{align}\label{eq:sph_t}
			0 = -\frac{1}{4} + \frac{1}{N} \sum_k \frac{\xi_k^2(t)}{2\lambda_k}
			\quad .
		\end{align}
                When the dynamics start from the equilibrium ground state at $t=0$, consistency with the equilibrium constraint \eqref{eq:sph} requires that $\lambda_k = \lim_{t\to 0}\omega_k(t) / g(t)$.

\section{Quench Dynamics}
\label{sec:results}
\begin{figure}	
\centering
\includegraphics[width=\linewidth]{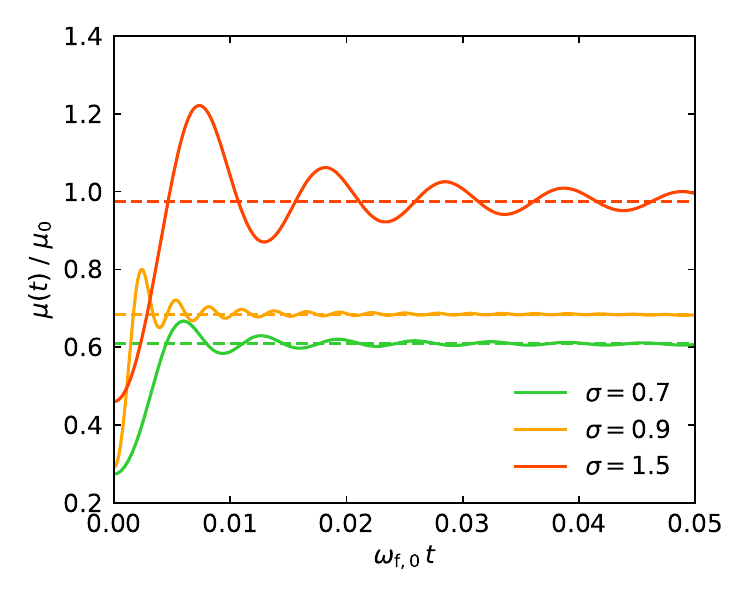}
	\caption{The time evolution of the Lagrange parameter $\mu(t)$ is shown following a quench from $g_0 = 2\gc$ to $\gf = \gc / 2$ in the spherical model on a finite one-dimensional lattice ($N=1000$) for different long-range interaction parameters $\sigma$. The system of equations \eqref{eq:erm} was integrated numerically, ensuring that the spherical constraint \eqref{eq:sph_t} be fulfilled to at least order $10^{-5}$ at all times. Dashed lines show the prediction for $\muf$ of the step approximation. Time is given in units of the post-quench frequency $\omfz$, calculated in the step approximation; $\mu(t)$ is shown in units of the initial Lagrange parameter $\mu_0$. Evidently, all the curves seem to converge to the predicted long-time values for quenches both into the gapped as well as into the gapless dynamical phase.}
\label{fig:mut}
\end{figure}
The dynamics under study is the one occurring after a sudden quench of $g$ at $t=0$, in other words
\begin{align}
g(t) = \begin{cases}
g_0  &\qfor* t \leq 0 \qc \\
\gf &\qfor* t > 0 . 
\end{cases}
\end{align}
In order to satisfy the constraint at all times, the Lagrange multiplier $\mu$ has to remain time-dependent also for $t>0$, where the coupling $g$ is constant. This essentially couples the equations for all $k$ modes and leads to a nontrivial time dependence of the frequency $\omega_k(t)$ and the solution of Eq.\,\eqref{eq:erm} cannot be obtained straightforwardly.

\subsection{Step approximation}
In order to simplify the solution of the problem, one may assume that as $g$ is quenched the Lagrange multiplier also discontinuously jumps from its initial value $\mu_0$ to another constant value $\muf$. The final value $\muf$ has to be chosen to coincide with the $t\to\infty$ limit of the time-dependent Lagrange multiplier, which is assumed to thermalize at  long times. A similar approximation has already been introduced for $O(n)$ models in the $n\to\infty$ limit and it is known as \emph{step approximation}\,\cite{sotiriadis_quantum_2010,chandran_equilibration_2013}.

Within this approximation, the frequency of the eigenmodes is simply quenched from $\omi$ to $\omf$ and the solution of Eq.\,\eqref{eq:erm} is given by
\begin{align}\label{eq:xi}
\xi_k(t) = \sqrt{1 + \epsilon_k \sin[2](\omf t)} 
\end{align}
with the parameter
\begin{align}
\epsilon_k = \left( \frac{\gf\,\omi}{g_0\, \omf} \right)^2 - 1 \quad .
\end{align}
We insert Eq.\,\eqref{eq:xi} into Eq.\,\eqref{eq:sph_t} and take the thermodynamic limit $N \rightarrow \infty$, i.e., turn the sum into an integral, which yields
\begin{align}
\label{td_const}
0 = \int\! \frac{\dd{k}}{2\pi} \frac{g_0}{2 \omi} \left[
\frac{\epsilon_k}{2} (1 - \cos 2\omf t) \right] \quad .
\end{align}
The time-dependent cosine term shows that a constant $\muf$ cannot satisfy the constraint \eqref{td_const} for short times after the quench. However, for long times the oscillating terms dephase for a continuum of $k$ modes, and the step approximation becomes exact. Therefore, it is sufficient to solve the Eq.\,\eqref{td_const} in the long-time limit, where the rotating wave approximation can be applied and the cosine term disregarded, yielding
\begin{align}\label{eq:eps}
0 = \int\! \frac{\dd{k}}{2\pi} \frac{\epsilon_k}{\omi}.
\end{align}
This implicit equation determines the long-time asymptotic value of $\muf$ via the $\mu$ dependence of $\omf$.

Therefore, the step approximation for $\mu$ is consistent with the long-time solution of the full constraint Eq.\,\eqref{eq:sph_t}. As a further proof of the applicability of the step approximation in the long-time limit, we solve the full differential equations set in Eq.\,\eqref{eq:erm}, fulfilling the constraint in Eq.\,\eqref{eq:sph_t} at each time step. 
In this way the Lagrange multiplier $\mu(t)$ can be obtained at each point $t$ in time. The results of this procedure are shown in Fig.\,\ref{fig:mut}  for a system of finite size $N=1000$ with quench parameters $g_0 = 2\gc$ to $\gf = \gc / 2$ at different $\sigma$ values. The resulting picture for the full dynamics is fully consistent with the equilibration behavior assumed in the step approximation. Indeed, after a short transient period, the curves $\mu(t)$  steadily oscillate around the predicted $\muf$ values and slowly converge to the solution of Eq.\,\eqref{eq:eps}. The aforementioned picture applies independently of the $\sigma$ value both in the gapped and gapless phases of Fig.~\ref{fig:dynphase}, as long as $\sigma>0$.

The full solution of the time-dependent constraint displayed in Fig.\,\ref{fig:mut} justifies the application of the step approximation in the following study of the dynamical phase transition. Accordingly, we are going to employ the step approximation to depict the entire dynamical phase diagram of the model, both in the region  $g_0 < \gc$ as well as for $g_0 > \gc$.

\subsection{Dynamical critical coupling}
Let us come back to Eq.\,\eqref{eq:eps} and discuss the conditions for the final Lagrange multiplier $\muf$ to attain the critical value $\muc$, at which the equilibrium would display the critical coupling $\gc$, see Eq.\,\eqref{eq:gc}. Using the same method,  one can define a dynamical critical coupling $\gcdyn$, such that for $\gf = \gcdyn$ the system becomes gapless and the constraint parameter $\muf$ in Eq.\,\eqref{eq:eps} approaches $\muc$. One, therefore, obtains the following equation for the dynamical critical coupling,
\begin{align}
\label{eq:gcdyn}
\half = \gcdyn \inv{\sqrt{g_0}} 
\idk \frac{\sqrt{2\mu_0 + U_k}}{2\muc + U_k},
\end{align}
which has been derived using the relation $2g_0 \idk (\omi)^{-1} = 1$, deduced from the equilibrium spherical constraint \eqref{eq:sph}.

\begin{figure}	
\centering
\includegraphics[width=\linewidth]{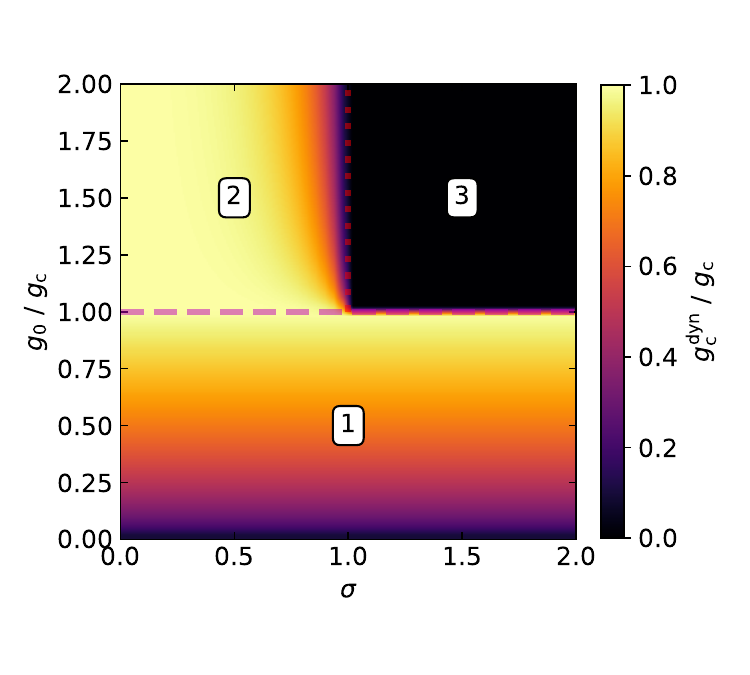}
\caption{The dynamical critical coupling $\gcdyn$ is shown as a function of $\sigma$ and initial coupling $g_0$ (in units of the equilibrium critical coupling $\gc$). The phase diagram contains three different regions: Region 1 is characterized by $g_0 \leq \gc$ where Eq.\,\eqref{eq:geomean} holds (colored region below red dashed line). In region 2 one has $g_0>\gc$ and $\sigma<1$, resulting in a finite $\gcdyn$, which decreases to zero as $\sigma$ tends to one. This leads to region 3 where $\sigma>1$ and $g_0>\gc$: in this region there is no ordered dynamical phase and $\gcdyn=0$ everywhere.} \label{fig:gcdyn}
\end{figure}

Based on Eq.\,\eqref{eq:gcdyn} one can identify two regimes: if $g_0 > \gc$, the integrand will diverge for each $\sigma \geq 1$, given that the denominator scales as $k^{\sigma}$, and no dynamical phase transition is present ($\gcdyn=0$). Instead, for $\sigma < 1$ the integral becomes convergent and it can be numerically evaluated to obtain the value of the dynamical critical coupling. If $g_0 \leq \gc$, on the other hand, the argument of the square root in the numerator of Eq.\,\eqref{eq:gcdyn} equals the denominator since $\mu_0=\muc$ in the whole low-temperature phase, leading to
\begin{align}\label{eq:gcdyn2}
	\half = \gcdyn \frac{1}{\sqrt{g_0}} 
		\idk \frac{1}{\sqrt{2\muc + U_k}} \quad ,
\end{align}
which is a rescaled version of the equilibrium relation in Eq.\,\eqref{eq:gc}. Accordingly, for $g_0 \leq \gc$, the critical value $\gcdyn$ is finite in the whole region $0<\sigma<2$, and is given by
\begin{align}\label{eq:geomean}
\gcdyn = \sqrt{g_0 \gc}
\end{align}
as shown in Fig.\,\ref{fig:gcdyn}.
		
\subsection{Loschmidt echo and correlation function}\label{sub:le}
\begin{figure*}
\centering
\includegraphics[width=0.9\linewidth]{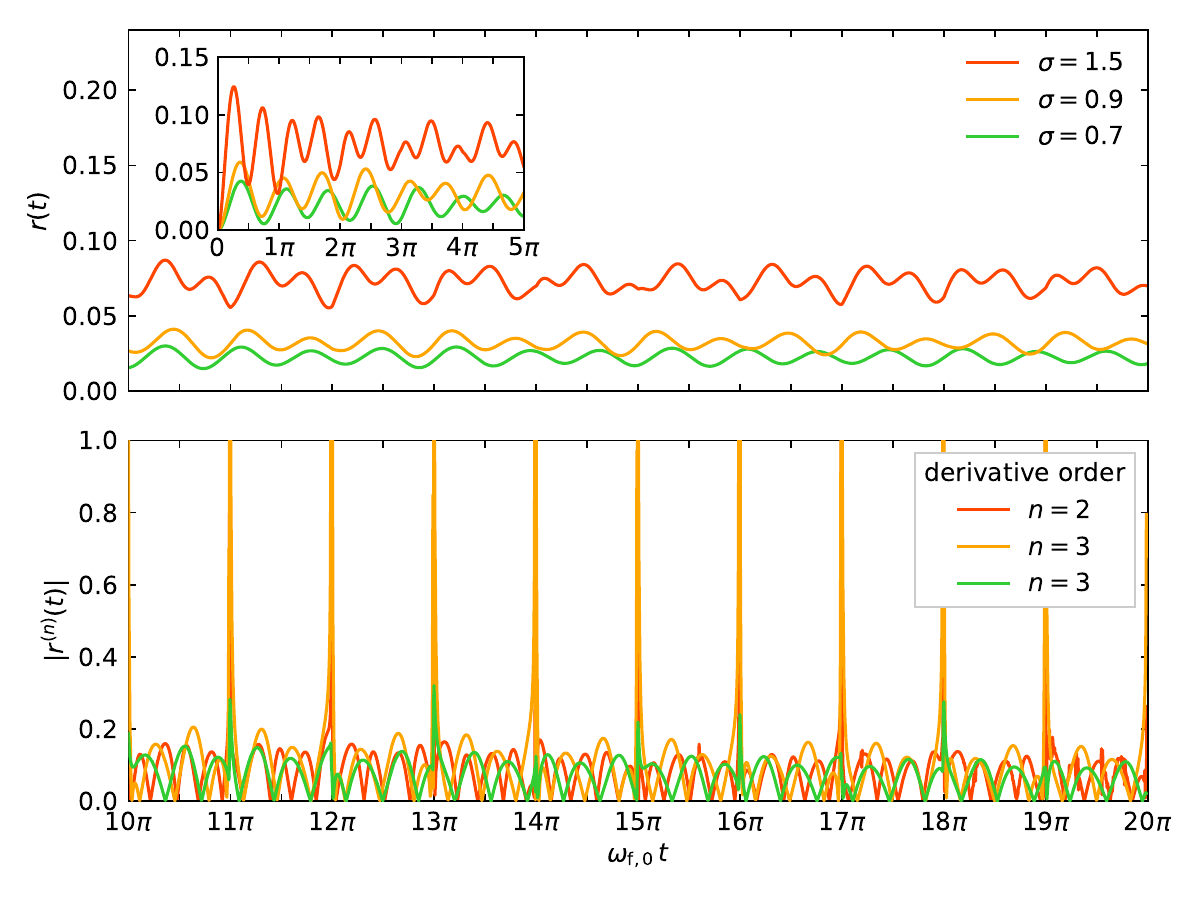}
\caption{Upper panel: The Loschmidt echo rate functions $r(t)$ for different $\sigma$ are shown following an order-to-disorder quench from $g_0 = \gc / \, 2$ to $\gf = 2\gc$. Because of the step approximation these are accurate for late times, hence we display the rate functions between the times $t^*_{10}$ and $t^*_{20}$. The inset shows the transient behavior of the same rate functions between $t^*_0=0$ and $t^*_5$ (the units on the axes are the same). Lower panel: the absolute values of the derivatives of the rate functions with respect to time are shown. At integer multiples of $\pi$ we observe divergences in the $n$-th derivative of the rate function whenever $n \sigma > 2$.} \label{fig:le}
\end{figure*}
In the previous section, we have depicted the appearance of the dynamical quantum phase transition, based on the vanishing of the gap in the single-particle spectrum. Such a dynamical quantum phase transition is also connected to the appearance of dynamical singularities in the Loschmidt echo rate function. In order to characterize such singularities, we will now focus on the dynamics of the system after a quench across the critical boundary $\gcdyn$. 

Let us consider an initial pure state in the ordered phase with vanishing initial gap $\Delta_0 = \mu_0 - \muc = 0$ and, suddenly, quench the system into the disordered finite-gap region $\Delta_f = \muf - \muc > 0$. In principle, a quench in the opposite direction from the gapped to the gapless phase may also be considered. However, in the latter case, the study is complicated by the divergence of the characteristic time scale of the system $1/\omf$, given by Eq.\,\eqref{eq:xi}, in the $k\to 0$ limit.
Indeed, the critical behavior is only influenced by the zero mode and, therefore, the time at which the first nonanalytic cusp appears is proportional to $1/\sqrt{\gap}$, which diverges for disorder-order quenches.

Thus, our focus will remain on order-to-disorder quenches with $g_0 < \gc$ and terminating in the gapped phase $\gf > \gcdyn$. When studying such dynamics, one shall consider that the spherical model will display a finite order parameter for $g_0 < \gc$, at least in the thermodynamic limit. The dynamics of the order parameter are coupled to that of the $k\neq0$ quantum modes, similarly to the case of the $O(n)$ models\,\cite{weidinger_dynamical_2017}. 
Equations of motion for the order parameter do not appear in our canonical transformation framework, but they can be obtained employing a time-dependent variational approximation\,\cite{cooper1986quantum},  which describes the dynamics of quadratic models exactly. However,  the resulting equations of motion  become numerically demanding and do not allow for the explicit introduction of the step approximation, so that it is more convenient to discard the order parameter contribution to the dynamics. Indeed, since our study will be carried out within the framework of the step approximation and the order parameter equilibrates to zero  in the long-time limit for $\gf > \gcdyn$, then the omission of the order parameter contributions is fully consistent with the following analysis.

Given the quadratic nature of the spherical model, one may calculate the overlap function 
\begin{align}
G(t) = \braket{\Psi_0(0)}{\Psi_0(t)} = \bra{0} \er^{-\ir Ht}\ket{0}
\end{align}
analytically. The representation of the time-dependent harmonic oscillator wave functions in position space reads
\begin{align}
\psi_{k,0}(x,t) =  \left( \frac{\omi}{\pi g_0 \xi_k^2(t)} \right)^{1/4} \er^{-\Omega_k(t) x^2 / 2 - \ir \varphi_k(t)},
\end{align}
where 
\begin{subequations}
\begin{align}
\Omega_k(t) &= \frac{\omi}{g_0 \xi_k^2(t)} - \ir \frac{\dot{\xi}_k(t)}{\gf \xi_k(t)}\;, \\
\varphi_k(t) &= \frac{\omi}{2g_0} \int_0^t \frac{\dd{t'} g(t')}{\xi_k^2(t')}\;.
\end{align}
\end{subequations}
The overlap is then given by 
\begin{align}
\begin{split}
	G(t) = &\prod_k \Biggl\{ \sqrt{2} \er^{-\ir\varphi_k(t)} \\ 
		&\times \biggl(  
				\xi_k(t) + \inv{\xi_k(t)} - 
				\ir \frac{g_0}{\gf} \frac{\dot{\xi}_k(t)}{\omi}  
				\biggr)^{\!\! -1/2} \Biggr\}  \ .
\end{split}
\end{align}
The Loschmidt echo rate function is obtained by the logarithm of the squared overlap
		\begin{align}
			r(t) &= -\lim_{N \rightarrow \infty} \frac{1}{N} 
				\log \abs{G(t)}^2  \\
			\begin{split}\label{eq:rate}
				&= -\log 2 + \int\! \frac{\dd{k}}{2\pi} 
					\log\abs{X_k(t)} \ ,	
			\end{split} 
		\end{align}
		where
		\begin{align}
			X_k(t) = \xi_k(t) + \inv{\xi_k(t)} - 
				\ir \frac{g_0}{\gf} \frac{\dot{\xi}_k(t)}{\omi} \ .
		\end{align}
		The integrand $X_k(t)$ in the expression for the rate function is a smooth function of time whenever $\omi$ is gapped.
		Hence, we should only expect to see nonanalytic cusps (and thus a dynamical quantum phase transition) for quenches starting in the gapless phase ($g_0 < \gc$).
		
		In the upper half of Fig.\,\ref{fig:le} we show the Loschmidt echo for a quench from $g_0 = \gc \, / \, 2$ to $\gf = 2\gc$ for different values of $\sigma$.
		At the critical times
		\begin{align}
			t^*_m = \frac{m \pi}{\omega_{\mathrm{f}, 0}} 
				\sim \frac{m}{\sqrt{\gap}}
				\qq{,} m \in \mathbb{N}
		\end{align}
		there are logarithmic divergences in the integrand in \eqref{eq:rate} which reflect as divergences in the derivatives of the rate function shown in the lower half of Fig.\,\ref{fig:le}.
		Since the critical time scale is set by the post-quench gap we do not expect to see nonanalytic cusps in the Loschmidt echo for a quench into the gapless phase as we also mentioned previously.

		The divergences show up in the $n$-th time derivative of the rate function whenever $n \sigma > 2$, due to the following reasons:
		For $k \rightarrow 0$ the function $X_k(t)$ diverges like $k^{-\sigma / 2}$ because of the term involving $1/\omi$. 
		However, at integer and half-integer multiples of $\pi$, the function $\dot{\xi}_0(t)$ is zero and cancels the divergence.
		Additionally, there is the term $1/\xi_k(t)$ which diverges only at half-integer multiples of $\pi$.
		All together, we find that $X_k(t)$ is divergent for $k\to0$ except at the critical times $t^*_m$.
		Differentiating \eqref{eq:rate} with respect to time $n$ times, we then encounter terms proportional to $(1/\omi)^n$ when $t = t^*_m$, which diverge like $k^{-n \sigma / 2}$. Since this is still integrated over $k$, the $n$-th derivative of the rate function diverges only if $n \sigma /2 > 1$, or $n \sigma > 2$.
		This analysis is valid in the whole region where $0<\sigma<2$.
		It does not apply for $\sigma \geq 2$ since there exists no gapless phase from which to start in that case.
		For $\sigma \leq 0$, an entirely different approach is needed due to energy extensivity breaking down.
		
The discussion above provides an additional evidence that the nonanalytic cusps are not merely a feature of the step approximation. Indeed, their emergence is a consequence of the initial conditions (i.e., starting in the gapless phase) and the particular form of the function $\xi_{k=0}(t)$ and its time derivatives, which will remain the same in the exact calculation. Moreover, the structure of the cusps remains unaltered in the long-time limit, where  $\mu(t)$ has equilibrated and the step approximation becomes exact.
		
\begin{figure}	
	\centering
	\includegraphics[width=\linewidth]{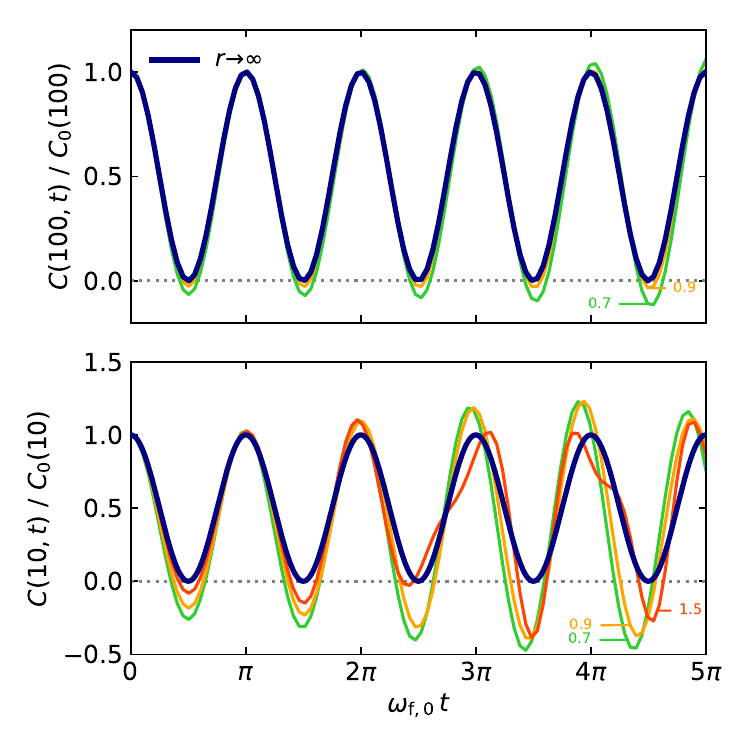}
	\caption{
		The correlation function $C(r,t)$ in units of the initial correlations $C_0(r)$ is shown as a function of time for different $r$ after a quench from the gapped to the gapless phase. 
		In the $r\to\infty$ limit Eq.\,\eqref{eq:sinecorr} is valid and the correlation function behaves like a simple sine curve independent of $\sigma$ (dark blue curve).
		For finite $r$ the correlations were numerically calculated using Eq.\,\eqref{eq:corr} in the large $N$ limit.
		They are clearly converging to the value of the large-$r$ approximation.
		Periodic dips and local maxima in the correlation function can be observed, lining up with the critical points at $t^*_m$.}
	\label{fig:corr}
\end{figure}
Following traditional results on spin systems\,\cite{zunkovic_dynamical_2018}, it would be interesting to connect the singular dynamics of the Loschmidt echo with the more traditional characterization of dynamical phase transitions based on order parameters. However, the contribution of the order parameter to the dynamics has been discarded in the present analysis and, in order to characterize the effect of the dynamical quantum phase transition on observables, we will follow a different route. 

A closed expression for the spatial correlation function can be obtained in terms of the dynamics of the quantum modes using the formula
\begin{align}\label{eq:corr}
	C(r, t) = \ev{s_i s_{j+r}} = \frac{g_0}{2N} \sum_k \frac{\xi_k^2(t)}{\omi} \er^{\ir k r}
\end{align}
by evaluating the sum at long distances $r \gg 1$.
In that limit we can employ the rotating wave approximation and see that the most important contribution to the sum arises from the values close to the infrared $k\to0$ limit.

We do this by splitting the integral into small and large wave-number contributions,
\begin{multline}
	C(r,t) = \half[g_0] \biggl[ \int_{-\Lambda}^{\Lambda} \dk 
		\frac{\xi_k^2(t)}{\omi} \er^{\ir k r} \\ +
		\int_{\Lambda \leq \abs{k} \leq \pi} \dk 
		\frac{\xi_k^2(t)}{\omi} \er^{\ir k r} \biggr] 
\end{multline}
with cutoff $0 < \Lambda \ll \pi$.
Utilizing the rotating wave approximation, it is evident that the second integral asymptotically approaches zero for large distances $r$ and does so faster than the first integral.
Since the dominant contributions to the integral then arise from a small shell around $k=0$, we can approximate the first integral by pulling $\xi_k^2(t)$ evaluated at $k=0$ outside the integral:
\begin{align}
	C(r,t) \simeq \xi_0^2(t) \half[g_0] 
		\int_{-\Lambda}^{\Lambda} \dk \frac{\er^{\ir k r} }{\omi} \ .
\end{align}
Note that for large $r$
\begin{align}
\label{op_fc}
	\half[g_0] \int_{-\Lambda}^{\Lambda} \dk \frac{\er^{\ir k r} }{\omi} 
	\simeq C(r,t=0) \equiv C_0(r) \ ,
\end{align}
such that
\begin{align}
	C(r,t) \simeq C_0(r) \xi_0^2(t) \ .
\end{align}
For a quench starting in the gapless phase, Eq.\,\eqref{eq:xi} yields
\begin{align}\label{eq:sinecorr}
	\frac{C(r,t)}{C_0(r)} \simeq \cos^2(\omfz t) \ .
\end{align}

When quenching from the gapless to the gapped phase in a DQPT, this leads to dips in the correlation function at half-integer multiples of $\pi$ as can be seen in Fig.\,\ref{fig:corr}.
At integer multiples of $\pi$, i.e., at the critical times $t^*_m$, we observe local maxima in the correlation function.

\section{Conclusion}\label{sec:end}
We have characterized the dynamical quantum phase transition occurring in the long-range interacting quantum spherical model by the study of the nonanalytic cusps in the Loschmidt echo following a quench of the coupling $g$. 
An approximate solution for the constrained system of differential equations governing the quench dynamics was developed by approximating the time-dependent Lagrange multiplier $\mu(t)$ as a step function in time.
A similar approximation was introduced by Ref.\,\cite{sotiriadis_quantum_2010} and employed by Ref.\,\cite{chandran_equilibration_2013} for the time-dependent effective mass of the $O(n)$ model in the $n\to\infty$ limit.
In the short-range limit $\sigma\to\infty$, our results for $\mu(t)$ are consistent with these previous works. In equilibrium the $n\to\infty$ limit of continuous $O(n)$ models lies in the same universality class as the lattice spherical model. Our result thus constitutes a further observation that the concept of universality may be extended to dynamical phase transitions. 
	
Using the step approximation we obtained an analytical expression for the dynamical critical coupling $\gcdyn$ as a function of initial coupling $g_0$ and the decay exponent $\sigma$.  The resulting phase diagram has been depicted in Fig.\,\ref{fig:gcdyn} and presents three different regions depending on the initial state of the system as well as on the value of the decay exponent $\sigma$. As already mentioned the short-range limit $\sigma\to\infty$ reproduces the results already found in Refs.\,\cite{sotiriadis_quantum_2010,chandran_equilibration_2013}.

Our model exhibits a dynamical quantum phase transition whenever we quench from the gapless equilibrium phase ($g_0 < \gc$) to the gapped dynamical phase ($\gf > \gc^{\mathrm{dyn}}$): then we encounter nonanalytic kinks in the Loschmidt echo rate function. Depending on $\sigma$ the smoothness of the rate function varies; for $\sigma > 1$ already the second time derivative of the rate function is discontinuous, while for lower values of $\sigma$ we have to go to higher and higher derivatives to see discontinuities.
The critical times with nonanalytic behavior are spaced apart by a critical time scale $T \sim \gap^{-1/2}$ inversely proportional to the post-quench gap. Because of this, nonanalytic cusps are not to be expected when quenching into the gapless phase due to the diverging time scale.

In equilibrium, the universal behavior of a long-range interacting system in dimension $d$ with decay exponent $\sigma$ can be related to the one of the corresponding nearest-neighbour system in dimension $d_{\mathrm{eff}}=\frac{2d}{\sigma}$\,\cite{joyce1966spherical,gori2017,defenu2020}. Applying the same relation to our findings, it is possible to reconstruct the dynamical phase diagram of the short-range interacting three-dimensional $O(n)$ model in the $n\to\infty$ limit\,\cite{weidinger_dynamical_2017}, showing that the foundations of universality also hold in the dynamical realm.
This can also be seen by noticing that the critical behavior only depends on the convergence properties of the integrals in Eqs.\,\eqref{eq:gc}, \eqref{eq:gcdyn2}, and \eqref{eq:rate} which in turn only depend on $d$ and $\sigma$.

However, it is worth noting that the relation between large-$n$ $O(n)$ models and the lattice spherical Hamiltonian in Eq.\,\eqref{eq:ham_eq} only holds for universal properties such as the shape of the dynamical phase diagram, but it does not imply a strict correspondence between the dynamical behavior of observables. Indeed, the connection between the Loschmidt echo and the correlation dynamics implies that local maxima of the correlations occur at the critical times where the rate function is nonanalytic. 
In other words, the rate function becomes nonanalytic when the time-evolved state recovers the magnetization profile of the initial state.
When the large-scale correlations are identified with the order parameter, this result differs from the one found in $O(n)$ models, where the zero crossings of the order parameter line up with kinks in the Loschmidt echo at the critical times\,\cite{weidinger_dynamical_2017}.

An interesting topic for future work would be to drive the coupling $g(t)$ slowly through the quantum phase transition instead of the instantaneous quench described here. Indeed, while the effect of long-range interactions on the universal dynamical scaling of critical Fermi systems has been investigated in detail\,\cite{defenu2019universal}, the effect of a slow drive in systems with bosonic excitations has mainly been discussed in the limiting $\sigma=-1$ case\,\cite{acevedo2014new,defenu2018dynamical}. The spherical model certainly offers a viable tool to investigate the adiabatic dynamics as a function of $\sigma$ in critical systems with bosonic excitations.

\begin{acknowledgments}
This work is supported by the Deutsche Forschungsgemeinschaft (DFG,
  German Research Foundation), project-ID 273811115 (SFB1225 ISOQUANT)
  and under Germany's Excellence Strategy EXC2181/1-390900948 (the
  Heidelberg STRUCTURES Excellence Cluster).
\end{acknowledgments}

\bibliography{references.bib}

\end{document}